\documentclass[%
reprint,
superscriptaddress,
showpacs,%
amsmath,amssymb,
aps,
prl,
floatfix,
]{revtex4-1}

\usepackage[utf8]{inputenc}
\usepackage[T1]{fontenc}
\usepackage[english]{babel}
\usepackage{amsfonts}
\usepackage{graphicx}
\usepackage{color}
\usepackage[normalem]{ulem}

\newcommand{\onefig}[1]{\centering{
    \includegraphics[width=0.99\columnwidth]{#1}}}
\newcommand{\widfig}[2]{\includegraphics[width=#1\columnwidth]{#2}}

\newcommand{\myii}{\mathrm{i}}
\newcommand{\mat}[1]{\mbox{\boldmath{$\mathrm{#1}$}}}
\newcommand{\determ}{\mathop{\text{det}}}
\renewcommand{\Re}{\mathop{\mathrm{Re}}\nolimits}
\renewcommand{\Im}{\mathop{\mathrm{Im}}\nolimits}
\newcommand{\const}{\mathrm{const}}

\begin{document}

\title{Hyperbolic Chaos of Turing Patterns}

\author{Pavel V. Kuptsov}\email[Electronic
address:]{p.kuptsov@rambler.ru}%
\affiliation{Department of Instrumentation Engineering, Saratov State
  Technical University, \\ Politekhnicheskaya 77, Saratov 410054,
  Russia}%
\author{Sergey P. Kuznetsov}%
\affiliation{Kotel’nikov’s Institute of Radio-Engineering and
  Electronics of RAS, Saratov Branch,\\ Zelenaya 38, Saratov 410019,
  Russia}%
\author{Arkady Pikovsky}%
\affiliation{Institute of Physics and Astronomy, University of
  Potsdam, Karl-Liebknecht-Str. 24/25, 14476 Potsdam-Golm, Germany}%

 \pacs{ 05.45.-a, 89.75.Kd, 05.45.Jn}


\date{\today}

\begin{abstract}
  We consider time evolution of Turing patterns in an extended system
  governed by an equation of the Swift-Hohenberg type, where due to an
  external periodic parameter modulation long-wave and short-wave
  patterns with length scales related as 1:3 emerge in succession.  We
  show theoretically and demonstrate numerically that the spatial
  phases of the patterns, being observed stroboscopically, are
  governed by an expanding circle map, so that the corresponding chaos
  of Turing patterns is hyperbolic, associated with a strange
  attractor of the Smale-Williams solenoid type.  This chaos is shown
  to be robust with respect to variations of parameters and boundary
  conditions.
\end{abstract}

\maketitle
In nonlinear dynamics the notion of structural stability, or
robustness, is one of the key tools allowing one to specify systems
and effects that are really significant for theoretical and numerical
researches, and especially for practical applications~\cite{Andronov,
  Shilnikov}.  Among chaotic attractors, structural stability is
intrinsic to those possessing the uniform hyperbolicity (``the systems
with axiom A''), mathematical examples of which were advanced already
since 60's -- 70's~\cite{Anosov,Smale,Williams,Plykin}. That time,
such attractors were expected to be relevant for various physical
situations (such as hydrodynamic turbulence), but later it became
clear that the chaotic attractors, which normally occur in
applications, do not relate to the class of structurally stable ones.
This is an obvious contradiction to the principle of significance of
the robust systems mentioned above.

Recently, this inconsistency has been partially resolved by
introducing a number of physically realizable systems with hyperbolic
chaotic attractors~\cite{Hyp,KuznetsovPikovsky,KuzBook,KuzUfn}. It has
been shown that simple systems of coupled oscillators that are excited
alternately (in time) possess hyperbolic attractors of Smale-Williams
type (for experimental realizations,
see~\cite{KuzUfn,KuzBook,HypExper}). Hyperbolic chaos in these systems
is related to the dynamics of the phases of the oscillators, evolution
of which on the successive stages of activity is governed by a
Bernoulli-type expanding circle map.

In this letter we develop a similar approach, but deal with \emph{the
  spatial phases} of patterns in a spatially extended system.  We
demonstrate the occurrence of hyperbolic chaos in dynamics resulting
from an interplay of two Turing patterns of different wave lengths
arising in succession.  This advance, first, extends a toolbox for
design of models manifesting robust chaos. Second, it suggests a novel
direction for search of situations associated with hyperbolic chaos in
the context e.g. of fluid turbulence, convection, and
reaction-diffusion systems. Third, the description in terms of
truncated equations for amplitudes of spatial modes leads to new
prototypical low-dimensional model systems with hyperbolic
attractors. (Note analogy with the Lorenz equations, which were
derived originally as a finite-dimensional model for fluid
convection.)

Let us illustrate the approach with a concrete example based on the
one-dimensional Swift-Hohenberg equation~\cite{Cross}. Consider its
following modification:
\begin{equation}
  \label{eq:swhhyp}
  \partial_{t} u+[1+\kappa^{2}(t)\partial_{x}^{2}]^{2} u=[A+B \chi(x)]u-u^{3}.
\end{equation}
Here $A$ is a positive parameter controlling the Turing
instability. An instant value of $\kappa$ determines the wave number
of the unstable Turing mode. In our case $\kappa (t)$ is assumed to be
a periodic function: $\kappa(t)=1$ for $nT\leq t<(n+1/2)T$, and
$\kappa(t)=1/3$ for $(n+1/2)T\leq t < (n+1)T$. This switching provides
the excitation of two distinct alternating in time Turing patterns
characterized by the dominating wave numbers, $k=1$ and $k=3$,
respectively.  The time interval $T$ between the switchings is
supposed to exceed the characteristic time duration of the formation
or decay of the Turing patterns.  A nonlinear cubic term in the
equation is responsible for saturation of the instability.  Also, the
coefficient at the linear term in the equation is assumed to depend on
the spatial coordinate that corresponds to the presence of a spatial
non-uniformity characterized by a function $\chi(x)$; its role will be
clarified below. Assuming the ring geometry and periodic boundary
conditions $u(x,t)\equiv u(x+L,t)$ (PBC), it is natural to set the
length of the system as $L=2\pi \ell$, with integer $\ell$, to get the
geometry supporting the Turing patterns of both the wave numbers $k=1$
and $3$.

The system operates as follows. In each time interval with
$\kappa(t)=1$ the Turing pattern with the dominating wave number $k=1$
arises, which is characterized by some spatial phase $\varphi$: $u\sim
U_1\cos (x+\varphi)+\tilde{U}_3\cos(3x+3\varphi)$, where
$\tilde{U}_3\ll U_1$, and $U_1$ is of the order of $\sqrt A$. (The
third harmonic appears naturally due to the cubic nonlinear term in
the equation.) After the switch to $\kappa=1/3$ the system becomes
unstable in respect to the harmonic component with $k=3$, while that
with $k=1$ starts to decay. The initial stimulation of the short-wave
pattern is provided by the component $\tilde{U}_3$; so, it accepts the
spatial phase $3\varphi$. At the end of the considered time interval
the first harmonic component practically disappears, and we have
$u\sim U_3\cos(3x+3\varphi)$, with $U_3$ of the order of $\sqrt A$.
After the next switch, when $\kappa=1$ again, the third harmonic
decays, but the first harmonic becomes unstable and starts to grow.  A
germ for this growth is provided by a component at the wave number
$k=1$ arising from the combination of the decaying short-wave pattern
and the spatially dependent coefficient $\chi(x)$.  If the Fourier
expansion of $\chi(x)$ contains a dominating second harmonic $k=2$,
the long-wave mode will arise with the phase $3\varphi$, due to the
term proportional to $\cos 2x
\cos(3x+3\varphi)=(1/2)\cos(x+3\varphi)+\ldots$.  Thus, on each
complete period of modulation $T$ the phase of the spatial pattern
undergoes the tripling (up to a constant phase shift): $\varphi_{n+1}
=3\varphi _{n}+\const$. This is an expanding circle map with chaotic
behavior characterized by the positive Lyapunov exponent $\Lambda =\ln
3\approx 1.0986$~\footnote{Also $\chi(x)$ can have the dominating
  fourth harmonic. In this case $\cos 4x
  \cos(3x+3\varphi)=(1/2)\cos(x-3\varphi)+\ldots$, so the map for the
  phase will be $\varphi_{n+1}=-3\varphi _{n}+\const$. Under the
  variation of relative weights of the components $k=2$ and $k=4$ some
  transitions between the topologically distinct behaviors will
  occur.}. Since the phase map is uniformly expanding, the
stroboscopic map corresponding to the transformation of the states
$u_n(x)\equiv u(x,t_n)$ from $t_n=nT+\const$ to $t_{n+1}$ is expected
to be hyperbolic.
\begin{figure}
  \onefig{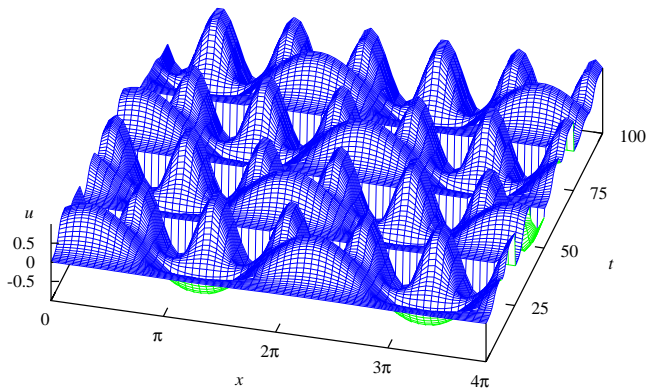}
  \caption{(Color online) Hyperbolically chaotic patterns in
    model~\eqref{eq:swhhyp} at $A=0.6$, $B=0.03$, $T=25$, $L=4\pi$,
    $N=64$, PBC.}
  \label{fig:sptm}
\end{figure}

Of course, this mode of operation occurs under the proper choice of
the parameters.  A value of $A$ is selected to get an instability at
$k=1$ with a decay at $k=3$, or vice versa, at successive half-periods
of parameter modulation.  The term $B\chi(x)$ must be small (comparing
to the fully developed pattern amplitude) to contribute only as a germ
for the formation of the long-wave pattern, although this germ should
be of a sufficient level to start the process with saturation on the
time scale $T$. In fact, the requirements are not very strict: the
described type of behavior occurs in a fairly wide parameter range.

\begin{figure}
  \centering\widfig{0.95}{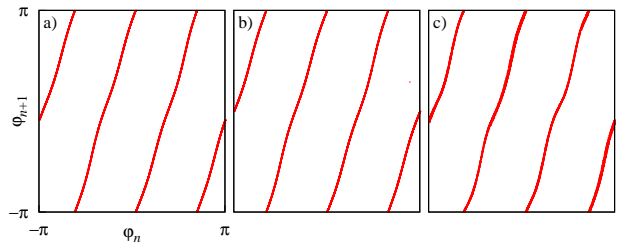}
  \caption{(Color online) Diagrams for spatial phases of Turing
    patterns at $t_n=(n+1/4)T$ for $A=0.6$, $B=0.03$, $T=25$. (a)
    Numerical solution of the system~\eqref{eq:swhhyp}, PBC, $L=4\pi$,
    $N=64$. (b) Amplitude equations~\eqref{eq:swhode}. (c)
    System~\eqref{eq:swhhyp} with ZBC, $L=8\pi$, $N=128$.}
  \label{fig:fmap}
\end{figure}
Figure~\ref{fig:sptm} illustrates the spatio-temporal behavior of the
system observed for the case of PBC. The 3D-plot $u(x,t)$ is obtained
using computations on a spatial grid with the node separation $\Delta
x=L/N$, where $N$ is a number of the nodes. One can observe the
alternating evolution of the Turing structures: a long-wave pattern
first appears, then decays, and is replaced by a short-wave one. After
the period $T$, the long-wave pattern reappears but with a different
spatial phase (shift along $x$-axis), and the process repeats.  As we
show in Fig.~\ref{fig:fmap}(a), the spatial phases recorded
stroboscopically follow a chaotic map of the expected type.  To obtain
this diagram, we determine the spatial phases at $t_n=(n+1/4)T$ as
$\varphi_n=\arg [u({L/2},t_n)+\myii\partial_xu({L/2},t_n)]$, where the
spatial derivative $\partial_xu$ is estimated by the numerical
differentiation, and the results are plotted in coordinates
$\varphi_{n+1}$ versus $\varphi_n$. This empirical map is of the
expected topological type: one revolution for the pre-image
corresponds to three revolutions for the image.

\begin{figure}
  \centering\widfig{0.8}{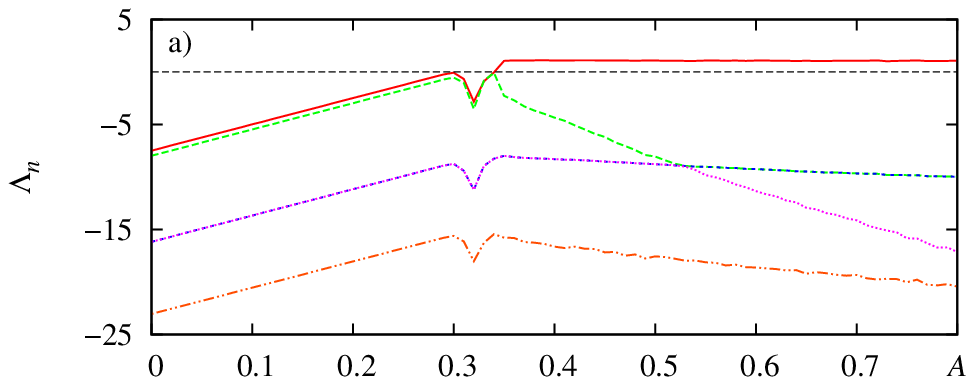}\\
  \centering\widfig{0.8}{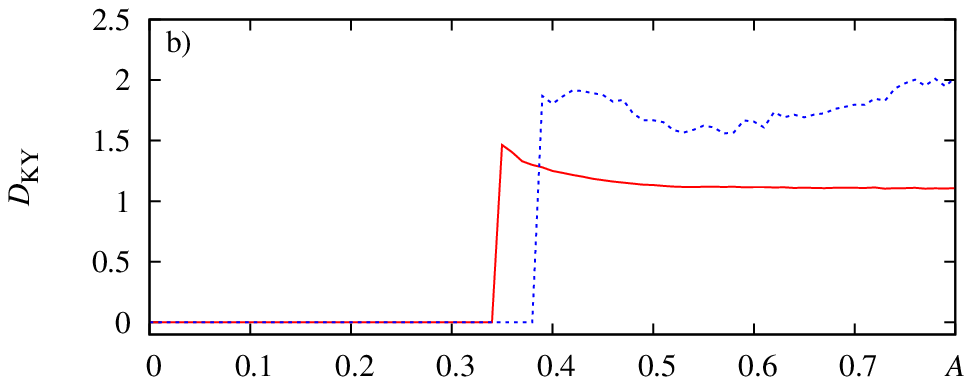}\\
  \centering\widfig{0.8}{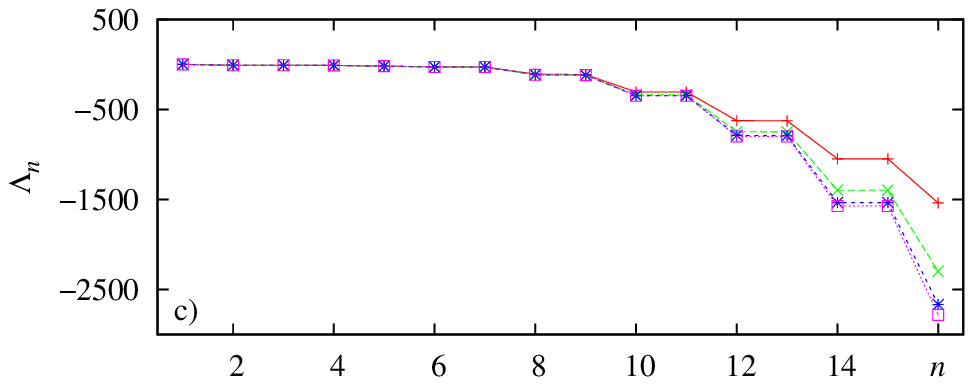}\\
  \centering\widfig{0.8}{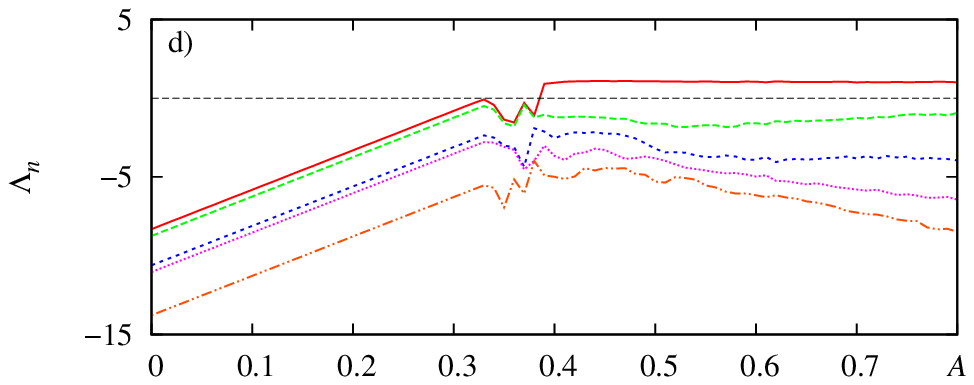}
  \caption{(Color online) (a) Five largest Lyapunov exponents vs. $A$
    for the stroboscopic map of the system~\eqref{eq:swhhyp} at
    $t_n=(n+1/4)T$, PBC. (b) Kaplan-Yorke dimension for PBC (solid
    line) and ZBC (dotted line). (c) PBC, first sixteen exponents for
    different $N$: pluses, crosses, stars and squares refer to $N=64$,
    $128$, $256$, and $512$, respectively. (d) ZBC, first five
    exponents. Other parameters are the same as in Fig.~\ref{fig:sptm}
    for PBC and $B=0.03$, $T=25$, $L=8\pi$, $N=128$ for ZBC.}
  \label{fig:lyap}
\end{figure}

To characterize chaos quantitatively and demonstrate its robustness,
we calculate the Lyapunov exponents.  Figure~\ref{fig:lyap}(a) shows
the first five Lyapunov exponents for the stroboscopic map as
functions of the parameter $A$.  The chaotic mode of operation occurs
above some threshold around $A \approx 0.38$.  In the chaotic regime
there is one positive Lyapunov exponent, which remains almost constant
in a wide parameter range. In particular, at $A=0.6$ the Lyapunov
exponents are $\Lambda=\{1.018, -9.34, -9.34, -11.42,-18.64, \ldots
\}$.  As expected, the largest exponent is close to $\ln 3$. As seen
from the diagram, all the exponents depend on the parameter smoothly,
without sharp spikes or dips. This is a manifestation of robustness of
the hyperbolic chaos~\cite{Hyp,KuzBook,KuzUfn}. The Kaplan-Yorke
dimension of the attractor varies slightly, see the solid line in
Fig.~\ref{fig:lyap}(b); in particular $D_\mathrm{KY}\approx 1.11$ at
$A=0.6$.

To confirm the validity of the used spatial discretization, in
Figure~\ref{fig:lyap}(c) we show the sixteen largest Lyapunov
exponents obtained at a fixed length $L$ with different sizes of the
numerical mesh $N$. The decrease of $\Delta x = L/N$ corresponds
obviously to approaching the continuous limit. The left-hand parts of
the curves overlap perfectly; so, the larger exponents are in good
correspondence for all tested step sizes.  The discrepancy visible in
the right-hand part of the plot for large negative exponents decreases
with the growth of $N$.  Hence, we can be sure that the properties
revealed in the computations with the finite discretization size are
valid for the continues system as well.

Next, we perform a direct test of the hyperbolicity. The hyperbolicity
implies that there are no tangencies between the stable and unstable
manifolds of orbits belonging to the attractor. Occurrence of a
tangency is determined by the zero angle between the expanding and
contracting tangent subspaces spanned by the corresponding covariant
Lyapunov vectors~\cite{GinCLV}. Following the method for testing
hyperbolicity described in~\cite{FastHyp12}, we examine the
distribution of these angles by considering the orthogonal complement
to the contracting subspace, which is normally much less dimensional
then the contracting subspace itself. If there are $K$ expanding
directions, it is sufficient to calculate $K$ orthogonal backward and
forward Lyapunov vectors, to construct a $K\times K$ matrix $\mat P$
of their scalar products, and to check how close to zero is the
normalized characteristic number
\begin{equation}
  \label{eq:detp}
  d_K=|\determ(\mat P)|.
\end{equation}
By the definition, $0\leq d_K\leq 1$. The procedure is applied at a
representative set of points on a trajectory on the attractor.  The
distribution of $d_K$ separated well from zero means that the chaos is
detected as hyperbolic: the tested trajectory does not contain any
points with tangencies of the expanding and conracting Lyapunov
vectors.

In application to the stroboscopic map of the system~\eqref{eq:swhhyp}
the calculations are simple because $K=1$.  For the parameters used in
Fig.~\ref{fig:sptm} we processed $10^5$ points and observed that
$(1-5\times 10^{-5})<d_1\leq 1$.  It means that the expanding
direction is always almost orthogonal to the contracting
subspace. Thus, the conjecture that the attractor is uniformly
hyperbolic is confirmed, but, of course, a rigorous mathematical proof
of the hyperbolicity would be desirable anyway.

As in the system only two modes with the wave numbers $k=1$ and $k=3$
are basically involved, one can expect that the essential properties
of the dynamics can be described with a truncated model.  To derive
the low-dimensional model we proceed as follows. Accounting for the
relevant modes, we use the ansatz $u=a_{1}(t)\cos x+b_{1}(t)\sin
x+a_{3}(t)\cos 3x+b_{3}(t)\sin 3x$. Assuming $\chi(x)=\cos 2x$, after
the substituting to Eq.~\eqref{eq:swhhyp}, we multiply the resulting
expression by $\cos x$ and $\sin x$, and by $\cos 3x$ and $\sin 3x$,
and for each case perform the integration over the spatial period
$2\pi$. The result is a set of equations for the amplitudes of the
modes, which can be compactly expressed in the complex form as
\begin{align}
    \dot{c}_1=&\mu_1 c_1-
    \textstyle{\frac{1}{4}}[3(|c_1|^2+2|c_3|^2)c_1-2Bc_3
    +(3c_1^*c_3-2B)c_1^*]\,,\nonumber \\
    \dot{c}_3=&\mu_3 c_3-
    \textstyle{\frac{1}{4}}[3(|c_3|^2+2|c_1|^2)c_3-2Bc_1+c_1^3]\,,
  \label{eq:swhode}
\end{align}
where the asterisk denotes complex conjugation, $c_1=a_1+ \myii b_1$,
$c_3=a_3+ \myii b_3$, $\mu_1=A-(1-\kappa^2)^2$,
$\mu_3=A-(1-9\kappa^2)^2$, and $\kappa=\kappa(t)$, as before.  Notice
that the structure of the equations resembles that for the amplitude
equations obtained for other models with hyperbolic attractors of
Smale-Williams type~\cite{KuzBook,KuznetsovPikovsky,Isaeva}.

\begin{figure}
  \centering\widfig{0.9}{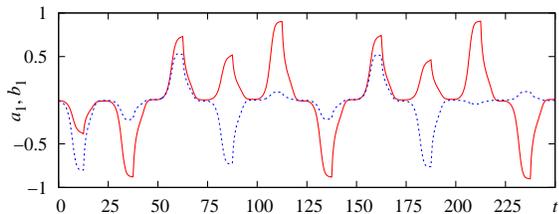}
  \caption{(Color online) Solution of Eq.~\eqref{eq:swhode} at
    $A=0.6$, $B=0.03$, $T=25$. Solid (red) and dotted (blue) lines
    refer to $a_1=\Re c_1$, and $b_1=\Im c_1$, respectively.}
  \label{fig:sptm_amp}
\end{figure}

Figure~\ref{fig:sptm_amp} illustrates the dynamics of the
model~\eqref{eq:swhode}. Observe the switchings after each next
half-period $T/2$. Different heights of the humps of $\Re c_1$ and
$\Im c_1$ arise due to the variations of the phases of $c_1$. The
phases transform stroboscopically according to the triple expanding
circle map; see Fig.~\ref{fig:fmap}(b) for the diagram for the phases
computed as $\varphi_n=\arg[c_1(t_n)]$. The Lyapunov exponents
evaluated for the stroboscopic map of the model~\eqref{eq:swhode} at
$A=0.6$, $B=0.03$, $T=25$ are $\Lambda=\{1.083, -12.5, -804.7,
-806.5\}$, and the Kaplan-Yorke dimension is $1.09$. Notice that the
first exponent is close to $\ln 3$. The hyperbolicity test described
above again shows that the expanding and contracting subspaces are
almost perfectly orthogonal.

Now we intend to demonstrate that the hyperbolic chaos can be observed
in geometries distinct from the ring one. In particular, let us
examine simple zero boundary conditions (ZBC): $u(x,t)=0$ for $x\leq
0$ and $x\geq L$, which could, in principle, block the mechanism of
the chaotic transformation of the phases (because near the ends the
spatial phase is dominated by the boundary conditions). However,
computations show that such a blocking occurs only in short
systems. If the length is large enough, patterns in the middle part of
the system still interact in the same way as for PBC, while the parts
close to the ends undergo deformations to fit ZBC; see
Fig.~\ref{fig:fmap}(c) and \ref{fig:lyap}(d) for $L=8\pi$. The map for
the spatial phases at the middle part of the system agrees well with
the expected form. Moreover, there is a large parameter interval,
where the system has a single positive Lyapunov exponent of value
almost independent on $A$. At $A=0.6$ the Lyapunov exponents are
$\Lambda=\{1.047, -1.59, -3.92, -4.97, -6.16, \ldots \}$, and the
Kaplan-Yorke dimension is $1.66$. The hyperbolicity test shows
pronounced separation of $d_1$ from the origin, although the
distribution is wider ($0.93<d_1<1$) than for PBC.

Summarizing, in this letter we have shown how the hyperbolic chaotic
dynamics can emerge in extended systems due to an interplay of spatial
patterns with different wave lengths. In our model system the spatial
phases of the patterns evolve in time according to the Bernoulli-type
tripling map, and their dynamics is strongly and robustly chaotic,
while the amplitudes behave in a rather regular manner. The mechanism
of the hyperbolic chaos is similar to that in alternately excited
oscillations, studied earlier~\cite{Hyp}.  In some respects, the
chaotization of spatial phases appears to be easier for implementation
(there is no necessity to have more than one involved subsystem). We
have demonstrated the expected chaotic behavior in the partial
differential equation of the Swift-Hohenberg type, and in the
truncated model represented by a set of ordinary differential
equations.  It should be emphasized that the kind of dynamics we
consider is not specific for the Swift-Hohenberg equation
only. Ingredients needed for the phase multiplication mechanism,
namely, the alternation of patterns due to parameter modulation, the
non-linearity, and the spatial inhomogeneity can be either found or
created in many spatially extended systems.  As expected, these
results open prospects for the search and constructing for hyperbolic
chaos in pattern-formation for systems in fluid dynamics (Faraday
ripples, convection rolls) and in reaction-diffusion systems (Turing
structures, advection induced patterns)~\cite{Cross}. In the case of
microfluidic systems~\cite{MicromixRev,Micromix}, an interesting
question for future studies is the effect of hyperbolic chaos on
Lagrangian mixing properties.

The authors acknowledge support from the RFBR grant No 11-02-91334 and
DFG grant No PI 220/14-1.

\bibliography{hyptur}

\end{document}